\documentclass[doublecol]{epl2} 

\title{Optimal transport on wireless networks}

\author{Y. Yu\inst{1} \and B. Danila\inst{1} \and J. A. Marsh\inst{2} \and K. E. Bassler\inst{1}}
\shortauthor{Y. Yu \etal}

\institute{                    
  \inst{1} Department of Physics, The University of Houston, Houston TX 77204-5005 \\
  \inst{2} Assured Information Security, Rome NY 13440
}

\pacs{89.75.Hc}{Networks and genealogical trees}
\pacs{89.20.Hh}{World Wide Web, Internet}
\pacs{89.75.Da}{Systems obeying scaling laws}
\pacs{05.60.-k}{Transport processes}

\abstract{
We present a study of the application of a variant of a recently introduced heuristic algorithm for the optimization of transport routes on complex networks to the problem of finding the optimal routes of communication between nodes on wireless networks. 
Our algorithm iteratively balances network traffic by minimizing the maximum node betweenness on the network.
The variant we consider specifically accounts for the broadcast restrictions imposed by wireless communication by using a different betweenness measure.
We compare the performance of our algorithm to two other known algorithms and find that our algorithm achieves the highest transport capacity both for minimum node degree geometric networks, which are directed geometric networks that model wireless communication networks, and for configuration model networks that are uncorrelated scale-free networks.
}

\begin{document}

\maketitle

\noindent The study of transport on complex networks has attracted a great deal of interest in recent years~\cite{NewmanSIAM,Guimera,Sreenivasan_arXiv,YanPRE,OurPRE1,OurChaos,Krause,
Krause1,Krause2,Krause3,Gupta1,Gupta2,Bettstetter,EcheniquePRE,EcheniqueEPL,ZhaoLaiParkYe,
ParkLaiZhaoYe,TB,TKBHK,OurPRE2,Korniss,Korniss1}.
One of the most important problems is to determine the routes between pairs of nodes
that optimize the efficiency of the transport.
Oftentimes, the routes that are used on networks are the so-called shortest-path routes, 
which are the routes with the minimum number of hops between any two nodes. 
As the volume of the transport increases, this approach leads to congestion or jamming of 
highly connected nodes on the network called hubs. 
Interest has developed in finding the routes that allow a given network to bear the highest 
possible amount of traffic without jamming~\cite{Guimera,Sreenivasan_arXiv,YanPRE,OurPRE1, OurChaos,Krause}.
This is done in general by defining the length of a path as a sum of weights assigned to the links that form it, and then reweighting various links to spread the traffic throughout the network and avoid jamming at hubs. 
The problem of finding the exact optimal routing has been shown~\cite{Sreenivasan_arXiv,Bui} to be $NP$-hard. 
Recently, however, we introduced a new heuristic routing optimization algorithm and showed~\cite{OurPRE1,OurChaos} that it finds near-optimal routing on both random (Erd\H{o}s-R\'enyi)~\cite{Erdos} and scale-free~\cite{BarabScience} networks.
Remarkably, this algorithm runs in only polynomial time $O(N^3\log N)$.
We also found that optimal routing preserves the small world character of networks~\cite{WattsStro}.

In this paper we use a variant of our algorithm to find optimal routes for transport on wireless communication networks, which has been the subject of a number of recent papers 
\cite{Krause,Krause1,Krause2,Krause3,Gupta1,Gupta2,Bettstetter}. 
Wireless networks are described by variants of random geometric networks. These networks lack the small world effect~\cite{WattsStro} that characterizes all of other types of networks we have previously applied our algorithm to. Transport on wireless networks occurs only along the subset of links that are bidirectional.
However, in order to avoid broadcasting interference, every time a node broadcasts an 
information packet the nodes at the ends of all of its outgoing links, whether they are bidirectional or not, are prevented from broadcasting or receiving packets.
The variant of our algorithm we consider here accounts for these broadcasting restrictions.

In order to study the effectiveness of our optimization of routing on wireless networks  and to understand how the optimization process works, we will compare results given by our optimal routing algorithm (OR) with those obtained using the shortest path routing algorithm (SP) and with the algorithm introduced in Ref.~\cite{Krause} (KR). 
The three algorithms will be applied to two different types of networks: minimum node degree geometric networks that, as stated above, are good models for wireless communication networks~\cite{Bettstetter}, and configuration model networks that are uncorrelated scale-free networks~\cite{Molloy}. Note that recent studies have shown that scale-free distributions of the node degrees are achievable~\cite{Rozenfeld,Herrmann,Duch} on geometric networks by a community-aware manipulation of the broadcasting powers of the nodes.

Minimum node degree geometric networks are constructed by uniformly distributing $N$ nodes randomly on a unit square and then adjusting their broadcasting powers until each node achieves a minimum degree $k_{min}$ counting only bidirectional links. They differ from random geometric networks~\cite{Dall} by their highly peaked degree distribution, with an average degree $\left< k\right>$ very close to $k_{min}$. Random geometric networks, by contrast, are obtained by connecting all pairs of nodes situated at a geometric distance shorter than a given threshold and are characterized by a binomial degree distribution. To facilitate comparison with Ref.~\cite{Krause}, the minimum degree $k_{min}$ is taken to be $8$.
Networks generated using the configuration model are characterized by a much broader power-law distribution of the node degrees, $p(k)\propto k^{-\gamma}$, and by the absence of any correlation between the degrees of adjacent nodes. 
The node degrees are allowed to vary between a lower cutoff $m$ and the square root of the number of nodes $N$. In our simulations we used $m=2$ and $\gamma=2.5$. 
For both models we considered networks with $N$ between 30 and 1600.
Both unconstrained routing as well as routing with wireless broadcasting constraints are considered for each case.

Our results show that the new variant of our algorithm achieves a significant improvement over the one presented in Ref.~\cite{Krause} for minimum node degree geometric networks. 
However, this improvement is not as large as we achieved for scale-free networks.
We will argue that this reduction in optimization efficiency is due to constraints on rerouting imposed by the non-small world nature of geometric networks.

Routing on the network is assumed to be done according to a static protocol which prescribes the next hop(s) for a packet currently at node $i$ and whose destination is node $t$. Each node is assumed to have a packet queue which works on a ``first-in/first-out" basis. When a new packet is added to the network at some node or arrives at a new node along its path, it is appended at the end of the queue. Upon reaching their destination, packets are removed from the network. For simplicity, we assume that all nodes have the same processing capacity of 1 packet per time step (assuming they are not inhibited by broadcasting neighbors) and that new packets are inserted at every node at the same average rate of $r$ packets per time step. This average insertion rate characterizes the load of the network. The destinations of the packets inserted at node $i$ are chosen at random from among the other $N-1$ nodes on the network. The algorithm can, however, be generalized for nodes with different processing capacities and for arbitrary traffic demands.

Given a loop-free routing table, the betweenness $b_i^{(s,t)}$ of node $i$ with respect to a source node $s$ and a destination node $t$ is defined~\cite{NewmanPRE} as the sum of the probabilities of all paths between $s$ and $t$ that pass through $i$. The total betweenness $B_i$ is found by summing up the contributions from all pairs of source and destination nodes. The practical way~\cite{NewmanPRE} to compute $b_i^{(s,t)}$ for all $i$ and $s$ is as follows: all nodes are assigned weight 1 and then the weight of every node along each path towards $t$ is split evenly among its predecessors in the routing table on the way from $t$ to $s$ and added to the weights of the predecessors.
\begin{figure}
	\scalebox{0.35}[0.35]{\includegraphics*{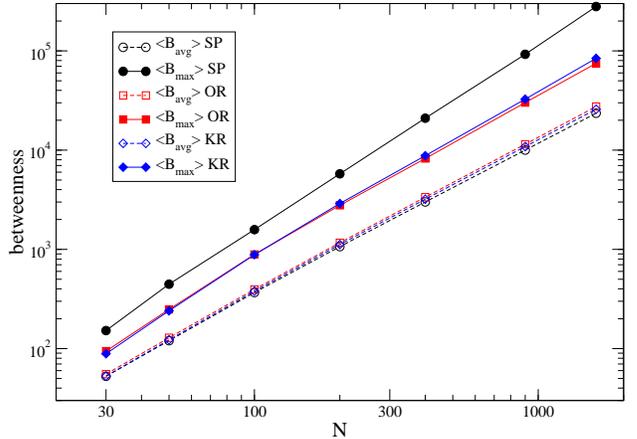}}
	\caption{(Color online) Ensemble averages of the average and maximum betweenness as functions of network size for minimum node degree geometric networks. Lower three sets (hollow black circles, red squares, and blue diamonds) represent $\left<B_{avg}\right>$, and upper three sets (solid black circles, red squares, and blue diamonds) represent $\left<B_{max}\right>$.}
    \label{fig.1}
\end{figure}
\begin{figure}
	\scalebox{0.35}[0.35]{\includegraphics*{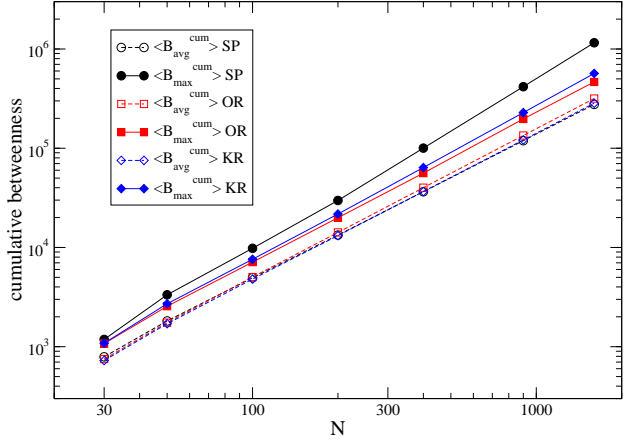}}
	\caption{(Color online) Ensemble averages of the average and maximum cumulative betweenness as functions of network size for minimum node degree geometric networks. Lower three sets (hollow black circles, red squares, and blue diamonds) represent $\left<B_{avg}^{cum}\right>$, and upper three sets (solid black circles, red squares, and blue diamonds) represent $\left<B_{max}^{cum}\right>$.}
    \label{fig.2}
\end{figure}

The aforementioned broadcasting restrictions are equivalent to saying that every node is processing not only the information packets passing through itself, but also those passing through its incoming neighbors. This situation can be accounted for by using the cumulative betweenness
\begin{equation}
\label{eq.1}
    B_i^{cum}=\sum_{k\in \mathcal{N}_i} B_k,
\end{equation}

\noindent where the incoming neighborhood $\mathcal{N}_i$ is the set formed by node $i$ together with all its incoming neighbors.

The time average of the number of packets passing through a given node $i$ in the course of a time step is
\begin{equation}
\label{eq.2}
    \left<w_i\right>_t=\frac{r B_i}{N-1},
\end{equation}

\noindent while the time average of the number of packets passing through its incoming neighborhood is
\begin{equation}
\label{eq.3}
    \left<w_i^{cum}\right>_t=\frac{r B^{cum}_i}{N-1}.
\end{equation}

\noindent Without broadcasting constraints, jamming of the network occurs at the critical average insertion rate $r_c$ at which the average number of packets processed by the busiest node reaches unity. Consequently, $r_c$ is given by~\cite{Guimera}
\begin{equation}
\label{eq.4}
	r_c=\frac{N-1}{B_{max}},
\end{equation}

\noindent where $B_{max}$ is the highest betweenness of any node on the network. If broadcasting constraints are considered, the critical insertion rate is determined by the busiest incoming neighborhood and we have
\begin{equation}
\label{eq.5}
    r_c^{cum}=(N-1)/B_{max}^{cum}.
\end{equation}

\noindent Thus, to achieve optimal routing on a wireless network, the highest cumulative betweenness $B^{cum}_{max}$ should be minimized~\cite{Krause2}.

The application of our algorithm to ordinary networks has been described in Ref.~\cite{OurPRE1}. For wireless networks, the algorithm proceeds as follows:

1. Assign uniform weight to every bidirectional link (SP routing) and compute the shortest paths between all pairs of nodes and the betweenness of every node.

2. Find the node $i_0$ which has the highest cumulative betweenness $B^{cum}_{max}$ and then the node with the highest ordinary betweenness in its incoming neighborhood $\mathcal{N}_{i_0}$. Increase the weight of every bidirectional link that connects the latter node to other nodes by adding half the initial weight to it.

3. Recompute the shortest paths and the betweennesses. Go back to step 2.

\noindent To achieve the $O(N^3\log N)$ running time, a binary heap must be used in the Dijkstra algorithm~\cite{Cormen} for the computation of the shortest paths to reduce the time required to sort the nodes by distance.

The algorithm described in Ref.~\cite{Krause} treats bidirectional links as two separate directed links and uses the cumulative betweenness of the node of origin as link weight. All cumulative betweennesses are initially set equal to 1 for the purpose of routing computation and then two rounds of iterations are performed. In the course of each round all shortest path routes are computed in the order of their node of origin, based on the latest values of the cumulative betweennesses. These values are updated after each computation of the shortest paths originating from a given node. This algorithm runs in time $O(N^2)$.

Throughout the paper, the network average of the betweenness $B_i$ is denoted by $B_{avg}$, while further averaging over an ensemble of network realizations is indicated by angular brackets. Fig.~\ref{fig.1} shows the ensemble averages of the network average and maximum betweenness, $\left<B_{avg}\right>$ and $\left<B_{max}\right>$ respectively, as functions of the number of nodes $N$ for minimum node degree geometric networks. Results are presented for shortest path routing (SP), for the optimal routing given by our algorithm (OR), and for the routing algorithm described in Ref.~\cite{Krause} (KR). The results for the average cumulative betweennesses $\left<B_{avg}^{cum}\right>$ and $\left<B_{max}^{cum}\right>$ for minimum node degree geometric networks are shown in fig.~\ref{fig.2}. For small networks of up to approximately 100 nodes the scaling of $\left<B_{avg}^{cum}\right>$ and $\left<B_{max}^{cum}\right>$ with network size is somewhat anomalous, but for larger networks both quantities scale with network size according to power laws. The exponents of the power laws were computed by fitting data corresponding to $N$ between 200 and 1600. The values of the exponents are given in table~\ref{tab.1}, with the quoted errors being $2\sigma$ estimates. It is apparent from figs.~\ref{fig.1} and~\ref{fig.2} and from table~\ref{tab.1} that our algorithm achieves the lowest maximum betweenness for a given network size. Both optimization algorithms obtain a significant improvement over shortest path routing. This is unlike the case of random or scale-free networks, where our algorithm brings a significant improvement in transport capacity when compared to any other optimization algorithm.
\begin{figure}
	\scalebox{0.35}[0.35]{\includegraphics*{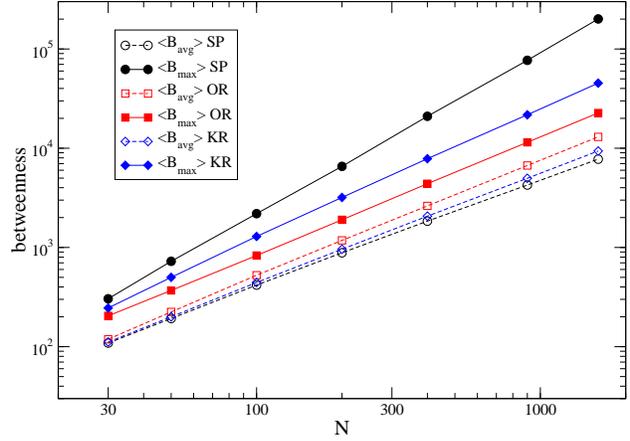}}
	\caption{(Color online) Ensemble averages of the average and maximum betweenness as functions of network size for uncorrelated scale-free networks. Lower three sets (hollow black circles, red squares, and blue diamonds) represent $\left<B_{avg}\right>$, and upper three sets (solid black circles, red squares, and blue diamonds) represent $\left<B_{max}\right>$.}
    \label{fig.3}
\end{figure}

There are two topological reasons for this less significant difference. The first reason is the lack of a small world effect~\cite{WattsStro}. In geometric networks with node-to-node communication ranges much smaller than the physical size of the network any shortest paths follow approximately the geometric shortest path and pass on average through a number of nodes proportional to the square root of the geometric distance between source and destination. On virtually all other types of networks of practical interest the average number of hops along the path increases with network size slower than logarithmically. This absence of shortcuts causes network traffic to be quite evenly spread even in the case of SP routing and reduces the likelihood of finding alternative paths that lower the maximum betweenness. Computation of $\left<L_{avg}\right>$ shows that, even in the case of the minimum node degree model where the broadcasting powers of the nodes are not equal, the average number of hops along the path scales with network size approximately as $\sqrt{N}$. The lack of a small world effect could be remedied by adding a few long range connections between randomly chosen pairs of nodes~\cite{Helmy,Lu}. This can be done in practice by scattering a few special nodes equipped with long range unidirectional antennas. The second reason for the less significant difference between the two algorithms is the highly peaked distribution of the node degrees which is a characteristic of the minimum node degree model and also contributes to the uniformity of traffic spreading.
\begin{table}
\label{tab.1}
\begin{center}
	\begin{tabular}{|c|c|c|c|}
	\hline
		  & SP & OR & KR \\
	\hline
    $\left<B_{avg}\right>$       & $1.488\pm 0.003$ & $1.519\pm 0.006$ & $1.509\pm 0.003$ \\
    $\left<B_{max}\right>$       & $1.874\pm 0.023$ & $1.585\pm 0.019$ & $1.619\pm 0.013$ \\
    $\left<B_{avg}^{cum}\right>$ & $1.459\pm 0.004$ & $1.492\pm 0.002$ & $1.477\pm 0.002$ \\
    $\left<B_{max}^{cum}\right>$ & $1.759\pm 0.005$ & $1.495\pm 0.005$ & $1.568\pm 0.007$ \\
	\hline
	\end{tabular}
	\caption{Exponents of the $\left<B_{avg}\right>$, $\left<B_{max}\right>$, $\left<B_{avg}^{cum}\right>$, and $\left<B_{max}^{cum}\right>$ power-law scaling with network size $N$ for minimum node degree geometric networks.}
\end{center}
\end{table}
\begin{table}
\label{tab.2}
\begin{center}
	\begin{tabular}{|c|c|c|c|}
	\hline
		  & SP & OR & KR \\
	\hline
    $\left<B_{max}\right>$       & $1.626\pm 0.011$ & $1.184\pm 0.012$ & $1.308\pm 0.010$ \\
    $\left<B_{max}^{cum}\right>$ & $1.799\pm 0.010$ & $1.480\pm 0.013$ & $1.617\pm 0.008$ \\
	\hline
	\end{tabular}
	\caption{Exponents of the $\left<B_{max}\right>$ and $\left<B_{max}^{cum}\right>$ power-law scaling with network size $N$ for uncorrelated scale-free networks.}
\end{center}
\end{table}

\begin{figure}
	\scalebox{0.35}[0.35]{\includegraphics*{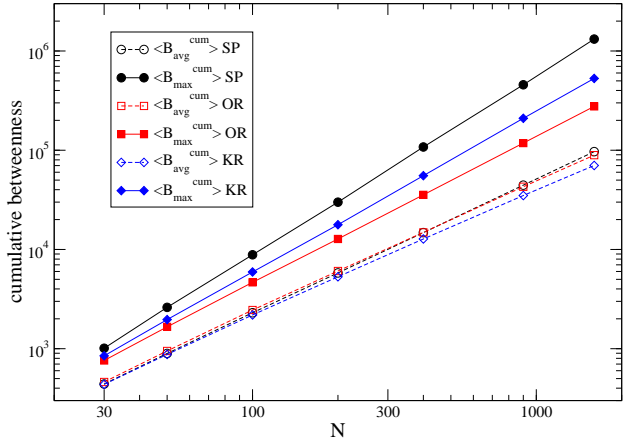}}
	\caption{(Color online) Ensemble averages of the average and maximum cumulative betweenness as functions of network size for uncorrelated scale-free networks. Lower three sets (hollow black circles, red squares, and blue diamonds) represent $\left<B_{avg}^{cum}\right>$, and upper three sets (solid black circles, red squares, and blue diamonds) represent $\left<B_{max}^{cum}\right>$.}
    \label{fig.4}
\end{figure}

Next, we computed $\left<B_{avg}\right>$, $\left<B_{max}\right>$, $\left<B_{avg}^{cum}\right>$, and $\left<B_{max}^{cum}\right>$ for uncorrelated scale-free networks generated using the configuration model using all three types of routing. Scale-free networks are characterized by an average number of hops which increases with the number of nodes slower than logarithmically. Results are shown in fig.~\ref{fig.3} for $\left<B_{avg}\right>$ and $\left<B_{max}\right>$ and in fig.~\ref{fig.4} for $\left<B_{avg}^{cum}\right>$ and $\left<B_{max}^{cum}\right>$. The exponents of the power laws for $\left<B_{max}\right>$ and $\left<B_{max}^{cum}\right>$ are given in table~\ref{tab.2}. The quantities $\left<B_{avg}\right>$ and $\left<B_{avg}^{cum}\right>$ vary in this case approximately as $N\log N$~\cite{OurChaos} and fitting them with a power law doesn't make sense from a theoretical point of view. The factor of improvement provided by our routing optimization algorithm is even more significant than in the case of the minimum node degree geometric network model and is comparable for both wireless and ordinary networks. This may become particularly useful if the geometric network topology of wireless networks is somehow altered (for example by introducing a community-aware distribution of the broadcasting powers) to resemble the topology of a small world network.

In summary, we have introduced a new algorithm for transport optimization on wireless networks and compared its performance with another recently introduced routing optimization algorithm. The effectiveness of transport optimization on wireless networks was compared to results obtained for ordinary networks with no broadcasting constraints. We found that our algorithm performs better in all cases studied, more significantly so in the case of scale-free networks. The less significant difference in the case of minimum node degree geometric networks is due to the geometric network topology (which results in a lack of a small-world effect) and to the narrowly peaked distribution of node degrees.

\acknowledgments
Support from the NSF through grant No.\ DMR-0427538 is acknowledged for Y.Y., B.D., and K.E.B. The authors thank Gy\H{o}rgy Korniss of Rensselaer Polytechnic Institute for useful discussions.

\end{document}